\documentclass[%
 aip,
 jmp,%
 amsmath,amssymb,
 reprint,%
]{revtex4-1}

\usepackage{graphicx}
\usepackage{dcolumn}
\usepackage{bm}
\usepackage{amssymb}   
\usepackage{times}

\begin{document}

\preprint{AIP/123-QED}

\title{Reflectionless design of optical elements using impedance-tunable coordinate transformation}

\author{Jun~Cao}
\affiliation{National Research Center for Optical Sensing/Communications Integrated Networking, Department of Electronics Engineering, Southeast University, Nanjing 210096, China }
\affiliation{Department of
Physics,
Nanjing Xiaozhuang University, Nanjing 211171,  China}
\author{Lifa~Zhang}%
\affiliation{Department of Physics, The University of Texas at Austin, Austin, Texas 78712, USA
}%
\author{Senlin~Yan}
\affiliation{Department of Physics, Nanjing Xiaozhuang University, Nanjing 211171,  China}
\author{Xiaohan~Sun}
\email{xhsun@seu.edu.cn}
\affiliation{National Research Center for Optical Sensing/Communications Integrated Networking, Department of Electronics Engineering,  Southeast University, Nanjing 210096, China }

\date{11 Feb 2014}

\begin{abstract}
We report a new strategy to remove the reflections resulted from
the finite embedded transformation-optical design by proposing
an impedance-tunable coordinate transformation, on which the functions
of impedance coefficients can be derived in the original space
without changing the refractive index. Based on the new approach,
two-dimensional (2D) reflectionless beam compressors/expanders, bends, shifters and splitters are designed using the modified anisotropic medium.
It is found that the reflections can be removed in magnetic response materials for TE polarization or dielectric response materials for TM polarization. The numerical simulations confirm that various reflectionless optical elements can be realized in the pure transformation optics. The impedance-tunable coordinate transformation can be generalized to three-dimensional (3D) cases and be applied to other transformation-optical designs.
\end{abstract}

\maketitle


 Transformation optics derived from form invariant coordinate transformations of Maxwell's equation, which has been reported by Pendry \cite{Pendry06} and Leonhardt \cite{Leonhardt06},  provides an
unconventional technique to design optical devices with excellent
functionalities, such as the invisible cloak \cite{Schurig06}, bends \cite{Huangfu08,Roberts08,Jiang08},
field rotators \cite{Chen07,Chen09}, concentrators \cite{Jiang08b,Yang09}, lens \cite{Schurig07,Kundtz10}.

The finite embedded coordinate transformation introduced by Rahm \cite{Rahm08b}
accelerates the transformation-optical design step, and provides
more flexibilities to the controlling and guiding of EM waves. However, such coordinate transformation is not a continuous
one where the impedance of the transformation
medium mismatches to those of the surroundings in many cases as reported in literatures \cite{Rahm08c,Jiang08c,Kwon08,Kwon09}. The most typical case is a 2D compressor/expander, suffering from the reflections inevitably as pointed in Ref.~\cite{Rahm08c}. Combining conventional method of inserting an antireflective coating, the reflections can be suppressed \cite{Garcia11}; but the coating is difficult to be realized in other cases with complicated boundaries. Emiroglu \cite{Emiroglu10} argued that
reflections can be removed under some 3D special
situations where the expanding/compressing rates are the same in two orthogonal directions. The reflections resulted from the discontinuous coordinate transformation limited its design in many applications. Therefore, a reflectionless transformation-optical design is still a open question up to now.

In this Letter, we propose an impedance-tunable coordinate transformation, based on which 2D reflectionless beam compressors/expanders, bends, shifters and splitters can be designed using anisotropic materials.

 In the traditional transformation-optical design \cite{Rahm08b,Rahm08c,Jiang08c,Kwon08,Kwon09,Garcia11,Emiroglu10}, the original space was set to be an invariable isotropic medium. We argue that the ratio of permittivity and permeability in the original space can be adjusted before transformation in order to compensate the ratio change induced by the asymmetric coordinate transformation, and make the transformation medium match the surroundings. We reset the permittivity $\varepsilon^{ij}=\varepsilon\delta^{ij}/k$
and the permeability $\mu^{ij}=k\mu\delta^{ij}$ in the original space, where the coefficient $k$ is spatial dependent, and is a continuous function to ensure the continuity of both original space and transformed space. For a given coordinate transformation $x^\prime=x^\prime(x)$, the permittivity $\varepsilon^{i^\prime j^\prime}$ and permeability $\mu^{i^\prime j^\prime}$ of the transformation medium calculated by the prescription of Ref. \cite{Schurig06b} can be expressed as \begin{eqnarray} \label{eq1}
\varepsilon^{i^\prime
j^\prime}=|det(A_i^{i^\prime})|^{-1}A_i^{i^\prime}A_j^{j^\prime}\varepsilon
\delta^{ij}/k,\nonumber\\
 \mu^{i^\prime
j^\prime}=|det(A_i^{i^\prime})|^{-1}A_i^{i^\prime}A_j^{j^\prime}k\mu\delta^{ij}.
\end{eqnarray}
where $A_i^{i^\prime}=\frac{\partial(x^\prime,y^\prime,z^\prime)}{\partial(x,y,z)}$
denote the Jacobian tensor between the transformed space
$(x^\prime,y^\prime,z^\prime)$ and the original space $(x,y,z)$.
Since in Eq.~(\ref{eq1}), the coefficient $k$  does not change the refractive index of the original space, as known in the geometrical optics \cite{Born99}, thus the light rays does not been changed both in original space and transformed space in comparison to the non-tunable case by the same coordinate transformation. However the ratio of the electric field
and the magnetic field have been changed by $k$, that is, the impedance coefficient is $k$ times as large as the original one. As we will see that pure transformation optics can manipulate the impedance match without combining conventional method of inserting an antireflective coating, nor the combining of field transformation approach \cite{Liu13}, which are need for bianisotropic medium. To obtain the reflectionless transformation medium, it is our task to figure out an appropriate function $k$.
\begin{figure}[t]
\includegraphics[width=3.40 in,  angle=0]{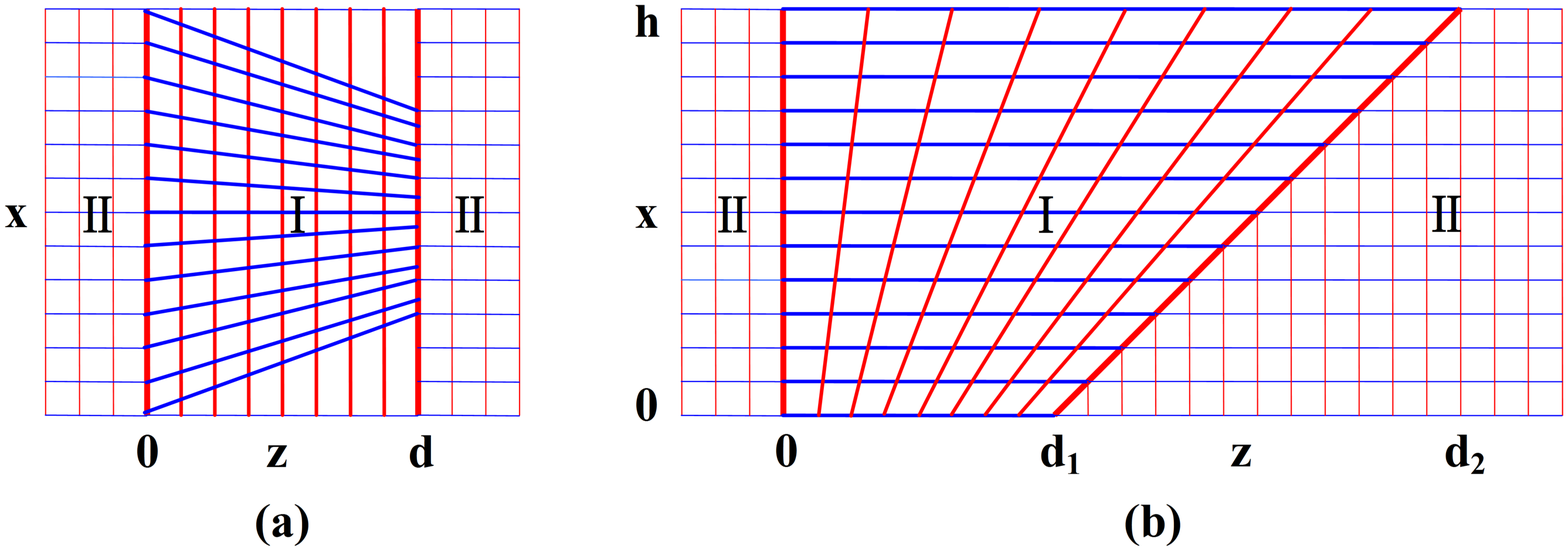}
\caption{ \label{fig1} (Color online)   Impedance-tunable coordinate transformation for the reflectionless design of (a) compressor; (b) bend. For both (a) and (b), the region I is transformation medium, the region II is air. }
\end{figure}

A 2D beam compressor of thickness $d$ has been designed as a first example, which structure is shown in Fig.~\ref{fig1} (a). To compress the wave propagating in the $+z$ direction without reflections, an impedance-tunable coordinate transformation has been done, which compresses the rectangular region in original space to trapezoidal region (the middle area 'I') in transformed space along $x$ axis linearly. The transformation
can be easily defined as $x^\prime=x[1-\frac{z}{d}(1-\gamma)]$,
$y^\prime=y$, $z^\prime=z$, where $\gamma$ is the compression
coefficient, $\gamma<1$ is for compressor and $\gamma>1$ for
expander respectively. For simplicity without loss of generality, the surrounding medium of the embedded transformation medium is supposed to be air, and the permittivity and permeability of the original medium is set to be $\varepsilon_0/k$ and $k\mu_0$ in all of our discussions.
Considering an incident plane wave upon the entrance boundary $z=0$ of
the compressor at an angle $\alpha$, it transmits from the exit
boundary $z=d$. One can easily obtain the reflectionless condition of $k = 1$ both for TE polarization ($E$-field along $y$ axis, associated with the parameters of $\varepsilon _{yy} $, $\mu _{xx} $, $\mu _{xz}  = \mu _{zx} $, and
$\mu _{zz} $) and TM polarization ($H$-field along y axis, associated with the parameters of $\mu _{yy} $, $\varepsilon _{xx} $,
$\varepsilon _{xz}  = \varepsilon _{zx} $, and $\varepsilon _{zz} $ ) plane waves at the entrance boundary $z=0$, where the transformation is continuous. In order to get the reflectionless condition, we need to calculate the reflection coefficient $R$ of the waves at the exit boundary to solve appropriate coefficient $k$. Applying the continuity of the total tangential electric and magnetic fields, at the exit boundary $z=d$ :
\begin{eqnarray}\label{eq_rzd}
R_{TE}  = \frac{{1 - k\gamma \sqrt {1 + \tan ^2 \alpha (1 - \frac{1}{{\gamma ^2 }})} }}{{1 + k\gamma \sqrt {1 + \tan ^2 \alpha (1 - \frac{1}{{\gamma ^2 }})} }}, \nonumber\\
R_{TM}  = \frac{{1 - \frac{\gamma }{k}\sqrt {1 + \tan ^2 \alpha (1 - \frac{1}{{\gamma ^2 }})} }}{{1 + \frac{\gamma }{k}\sqrt {1 + \tan ^2 \alpha (1 - \frac{1}{{\gamma ^2 }})} }}.
\end{eqnarray}
From Eq.~(\ref{eq_rzd}), one can obtain
$k = \frac{1}{{\gamma \sqrt {1 + \tan ^2 \alpha (1 - \frac{1}{{\gamma ^2 }})} }}$ for $R_{TE}=0$ and
$k = \gamma \sqrt {1 + \tan ^2 \alpha (1 - \frac{1}{{\gamma ^2 }})} $
for $R_{TM}  = 0$ at the exit boundary. To satisfy the reflectionless condition at both boundaries at the same time, and the continuity of the transformation medium, impedance coefficient $k$ should be spatial continuous function, and can be set as :
\begin{eqnarray}\label{eq_kc}
k|_{TE} =\frac{d}{{d - z'(1 - \gamma \sqrt {1 + \tan ^2 \alpha (1 - \frac{1}{{\gamma ^2 }})} )}}, \\
k|_{TM} =\frac{{d - z'(1 - \gamma \sqrt {1 + \tan ^2 \alpha (1 - \frac{1}{{\gamma ^2 }})} )}}{d}.
\end{eqnarray}
The function of $k$ is not unique, but should be easy for the realization of the transformation medium.

Note that for a vertical illumination ($\alpha = 0$), which is the general case for many applications, our selection of function $k$ can make the parameter of relative permittivity be $\varepsilon _{yy}  = 1$ for TE polarization, which is only a magnetic response. Therefore one only need to tune the relative permeability while keep the permittivity unchanged, which can be easily realized in metamaterials using split-ring resonators(SRRs) \cite{Schurig06}. Similarly, a dielectric response material ($\mu _{yy}  = 1$ ) can also be obtained for TM polarization by the above function $k$, which can be realized by metal wires \cite{Cai07}. The designed impedance-tunable compressor for vertical illumination can be transplanted to mode coupling in waveguide. Although some works \cite{Tichit10,Xu11} have been done in the metallic waveguide using the traditionally coordinate transformation, our approach can improve the performance with a supper-high efficiency. The method can be further applied to remove reflection resulted from mode mismatch in dielectric waveguide, which is a vital problem highly desirable to be solved in fiber-to-chip coupling.

Beside a compressor, a bend is also one of the most important elements in the optical design. Using the strategy of impedance-tunable coordinate transformation, we propose a compact 2D beam bend design, as shown in Fig.~\ref{fig1}(b). The bend transforms a rectangular region with a height of $h$ and a length of $d$
in the original space to trapezoidal region (the middle area 'I') with the unchanging height $h$ and the different length of $d_1$ and $d_2$ in the transformed space. The transformation is linear along $z$ axis and can be expressed by $x' = x$, $y' = y$, $z' = \frac{z}{d}(\frac{{d_2  - d_1 }}{h}x + d_1 )$. When a plane wave propagates through the transformed area along $+z$ axes, the equiphase surface will smoothly deflect an angle of $\beta  = \tan ^{ - 1} (\frac{{d_2  - d_1 }}{h})$, and the ray will be refracted to the air with the bending angle of $\beta$. For simplicity, a normal incident wave illuminating upon the bend is discussed here. Similar to the derivation of the compressor, the spatial continuous function $k$ for the reflectionless bend can be set as :
\begin{eqnarray}\label{eq_kb}
k|_{TE} = 1 + \frac{{z'}}{{d_1  + x'\tan \beta }}(\cos\beta  - 1),\\
k|_{TM} = 1 + \frac{{z'}}{{d_1  + x'\tan \beta }}(\frac{1}{{\cos\beta }} - 1).
\end{eqnarray}
Note that if only a single impedance-tunable bend applied, the bending angle
$\beta$ is limited from $0$ to $90^\circ$, and the size of the exit boundary of the bend is $\frac{1}{{\cos \beta }}$ times as large as that of the entrance boundary. The waves not only be bent but also be expanded with the decreasing magnitude due to the conservation of power, thus the magnitude of the bending waves is dependent on the bending angles. Two impedance-tunable bends in series will expand the range of the bending angles from $0$ to $180^\circ$ and controlling the magnitude of the bending waves independently.

\begin{figure}[t]
\includegraphics[ width=3.40 in,  angle=0]{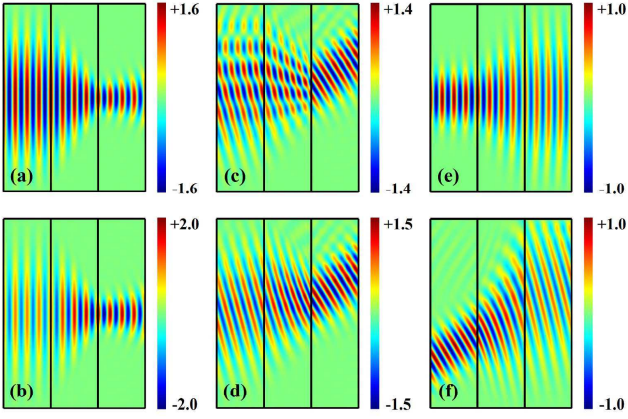}
\caption{ \label{fig2} (Color online) The normalized $y$ direction $E$-field distribution of a beam compressor and a beam expander. (a) the impedance-nontunable case and (b) the impedance-tunable case for the normal incidence plane wave upon the compressor while$\gamma  = 0.25$, (c)the impedance-nontunable case and (d) the impedance-tunable case for the oblique incidence plane wave with the angle $\alpha  = 15^\circ$ upon the compressor while $\gamma  = 0.5$, (e) and (f) are the impedance-tunable case of an expander with the same $k$  while $\gamma  = 2$ for $\alpha  = 0 $ and $\alpha  = 30^\circ $ respectively.}
\end{figure}

\begin{figure}[t]
\includegraphics[width=3.40 in,  angle=0]{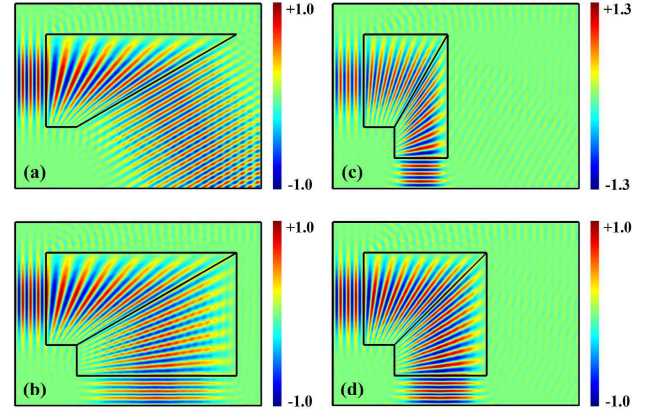}
\caption{ \label{fig3} (Color online)  The normalized $y$ direction $E$-field distribution of an impedance-tunable beam bend for normally-incident plane wave. (a) a single bend with the bending angle of $60^\circ$, two bends connected in series with a angle of $90^\circ$ with two bending angles $\beta _1  = 60^\circ,\, \beta _2  = 30^\circ $ for (b), $\beta _1  = 30^\circ,\,\beta _2  = 60^\circ $ for (c), and $\beta _1  = 45 ^\circ,\beta _2  = 45^\circ$ for (d).}
\end{figure}

To validate our theoretical results, 2D numerical simulations using COMSOL Multiphysics are carried out to investigate the performance of the TE polarization plane waves (similar calculation for TM waves can also be done) illuminating upon the compressors/expanders (Fig.~\ref{fig2}) and bends (Fig.~\ref{fig3}) respectively. The calculation domain is bounded by perfectly matched layers.  Fig.~\ref{fig2}(a) and Fig.~\ref{fig2}(b) plots the magnitude distributions of the electric field of an impedance-nontunable compressor and impedance-tunable compressor respectively with the compression coefficient $\gamma = 0.25$ to a normal incidence. When the original space is set to be invariable vacuum, as shown in Fig.~\ref{fig2}(a), reflections occur at the exit boundary of the compressor, resulting in an amplitude modulation of the incoming wave, and the simulated transmission efficiency is only $64\% $. Using impedance-tunable transformation by selecting the appropriate impedance coefficient $k$, as shown in Fig.~\ref{fig2}(b), reflections have been suppressed with the simulated transmission efficiency near $100\% $. The magnitude distributions of the electric field doubles at the exit boundary of the compressor compared to that at the entrance boundary. An oblique illumination upon the impedance-nontunable compressor and the impedance-tunable compressor with $\alpha  = 15^\circ$ are illustrated in Fig.~\ref{fig2}(c) and Fig.~\ref{fig2}(d). Note that a large compression rate ($\gamma  \ll 1$) will lead to the total reflection even for moderate incidence angle of $\alpha$, so we let $\gamma  = 0.5$. Fig.~\ref{fig2}(c) shows the distorted wave profile resulting from the reflections at the boundary of the impedance-nontunable compressor. While the wave profile is undisturbed in the impedance-tunable compressor as shown in Fig.~\ref{fig2}(d), with all the energy transmitted.
 Different impedance coefficient $k$  must be chosen to remove the reflections for different incident angle $\alpha$, while for the expander ($\gamma > 1$), $k$ changes slowly with big range of $\alpha $ even for an moderate expanding $\gamma  > 1$, which can be seen from Eq.~(\ref{eq_kc}). Fig.~\ref{fig2}(e) and Fig.~\ref{fig2}(f) show good performance of the expanding waves for $\gamma  = 2$ in the impedance-tunable expander with the same $k = 1 + \frac{{z'}}{d}(\frac{1}{\gamma }-1)$ for
$\alpha  = 0 $ and $\alpha  = 30^\circ$ respectively.

Similar simulations of the impedance-tunable beam bends are performed for a normal TE polarization plane wave, as shown in Fig. \ref{fig3}. Selecting the appropriate impedance coefficient $k$, no reflections occur and  all the transmission efficiencies are near $100\% $.  Fig.~\ref{fig3}(a) is a single bend with a bending angle of $60^\circ$; Fig.~\ref{fig3}(b), Fig.~\ref{fig3}(c) and Fig.~\ref{fig3}(d) are the cases for the bending angle of $90^\circ$ of two bends connected in series. The bending waves are expanded in Fig.~\ref{fig3}(b) and compressed in Fig.~\ref{fig3}(c). Combining two symmetrical bends with the same bending angle in Fig.~\ref{fig3}(d) where the transformation is continuous at the boundary, which is a special case designed in Ref. \cite{Huangfu08}, the impedance coefficient $k=1$ (the original space is a vacuum) can satisfy the reflectionless condition. Introducing a tunable impedance, we can design reflectionless bends where direction and magnitude of the waves are tuned independently, while it can not be realized in traditional transformation optics. Therefore impedance-tunable coordinate transformation provides much more convenience in general applications of optical bends design.

\begin{figure}[t]
\includegraphics[ width=3.40 in,  angle=0]{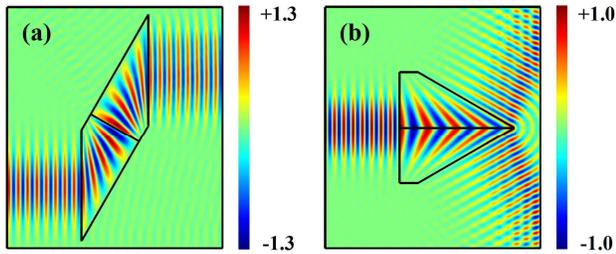}
\caption{ \label{fig4} (Color online) The normalized $y$  direction $E$-field distribution of a beam shifter (a) and a beam splitter (b) for normally-incident plane wave.  }
\end{figure}

As a further step, we can realize a reflectionless beam shifter by connecting  the two impedance-tunable bends with the opposite bending direction and keeping the entrance boundary and the exit boundary in parallel. As shown in  Fig.~\ref{fig4} (a), despite the discontinuity in both the entrance boundary and the exit boundary the transformation-optical design for the shifter is easily performed as well. A reflectionless beam splitter can also be realized by connecting the two impedance-tunable bends in parallel, as shown in Fig.~\ref{fig4} (b), Combining multi impedance-tunable bends, a multi road beam reflectionless splitter can be designed with the arbitrary beam angles and energy ratios or magnitudes.

In summary, to remove the reflections at the boundary of the transformation medium, we generalize the transformation optics by introducing a tunable impedance, in which a appropriate impedance coefficient are chosen in the original space. The impedance-tunable coordinate transformation is applied to the design of 2D compressors/expanders, the bends, the shifters and the splitters. The numerical simulation confirmed our design without any reflections. The proposed method provides more flexibilities to the controlling of EM waves, and can also be applied  to other  transformation-optical designs including 3D cases in the future.

This work was supported from the Frontier Research
and Development Projects of Jiangsu Province, China, under grants of BY2011147 and BY2013073.

\end{document}